# Can Large Language Models Become Policy Refinement Partners? Evidence from China's Social Security Studies

KE Jinghan[1*], ZHOU Zheng[2], ZHAO Yuxuan[1]

[1]*College of Humanities and Development Studies, China Agricultural University, China*

[2] *Baichuan Inc., China*

**Abstract**

The rapid development of large language models (LLMs) is reshaping operational paradigms across multidisciplinary domains. LLMs' emergent capability to synthesize policy-relevant insights across disciplinary boundaries suggests potential as decision-support tools. However, their actual performance and suitability as policy refinement partners still require verification through rigorous and systematic evaluations. Our study employs the context-embedded generation-adaptation framework to conduct a tripartite comparison among the American GPT-4o, the Chinese DeepSeek-R1 and human researchers, investigating the capability boundaries and performance characteristics of LLMs in generating policy recommendations for China's social security issues. This study demonstrates that while LLMs exhibit distinct advantages in systematic policy design, they face significant limitations in addressing complex social dynamics, balancing stakeholder interests, and controlling fiscal risks within the social security domain. Furthermore, DeepSeek-R1 demonstrates superior performance to GPT-4o across all evaluation dimensions in policy recommendation generation, illustrating the potential of localized training to improve contextual alignment. These findings suggest that regionally-adapted LLMs can function as supplementary tools for generating diverse policy alternatives informed by domain-specific social insights. Nevertheless, the formulation of policy refinement requires integration with human researchers' expertise, which remains critical for interpreting institutional frameworks, cultural norms, and value systems.

## 1 Introduction

The rapid advancement of artificial intelligence (AI) has propelled Large Language Models (LLMs) from laboratory-scale technical research to diverse real-world applications. Both academia and industry are actively exploring LLMs' capabilities in knowledge synthesis and scenario-specific deployment[1] LLMs now support mission-critical tasks in multiple sectors: in healthcare, they now assist clinicians in medical diagnostics with accuracy rates approaching or surpassing human experts[2]; in public administration, some governments have developed "AI civil servants" powered by LLMs, achieving significant improvements in document format correction and cross-departmental task allocation efficiency[3]; and in public opinion management, LLMs are used for text analysis of online discourse to identify and classify viewpoints and emotional expressions[4]. These implementations predominantly involve structured or semi-structured data: medical diagnostics integrate highly structured lab reports, medical histories, and imaging records with unstructured patient narratives; public administration relies on standardized document formats and workflow rules; and online discourse analysis utilizes highly structured semantic classification frameworks. However, their performance in processing unstructured data—particularly for synthesizing policy recommendations from raw research inputs—lacks scholarly examination.

---

* Corresponding author.

Formulating policy recommendations in social science represents a critical interdisciplinary challenge. As a nexus between academic insights and governance practices, it bridges theoretical depth and real-world complexity. Robust policy recommendations necessitate researchers' dual mastery of domain-specific social complexities and extant policy architectures. While there are widely accepted principles for policy design—such as resource concentration, clarity in implementation, flexible adaptability, challenging goals, integrated coordination, and internal and external consistency—policy recommendations in the social sciences are often highly context-dependent. These recommendations blend objective insights with subjective analyses, grounded in detailed descriptions and explanations of social issues. Ideal policy proposals should account for multiple dimensions: dynamic stakeholder interactions, historical trends in multisource data, intergenerational equity in policy tools, and long-term impacts of interventions, all of which pose significant challenges to researchers[5]. Just as medical diagnostics and therapeutics address an individual organism, policy recommendations represent the "diagnostic and therapeutic interventions" for a social organism. If LLMs can have applicable policy recommendation capabilities in the domain of social sciences, just as they have demonstrated the ability to provide diagnostic advice in the medical field, it would be of great assistance to researchers. However, unlike STEM fields (Science, Technology, Engineering, and Mathematics), policy recommendations in the social sciences are not merely technical issues. They also involve complex factors such as cultural ethics, values, and social consensus. The policy benefits are directly related to national governance effectiveness and societal welfare, thus requiring more rigorous validation and deliberation.

From a theoretical perspective, LLMs have certain potential advantages in proposing policy recommendations for social sciences[6]. Firstly, LLMs possess the ability to efficiently process large-scale text data, enabling them to quickly extract key information and identify potential patterns. This characteristic allows them to rapidly read a vast amount of descriptive factual materials and distill the key points, thereby providing a reliable information foundation for policy recommendations. Secondly, LLMs' depth of knowledge in specific domains and their ability to integrate cross-disciplinary knowledge even surpass that of human experts[7]. Training on a vast amount of cross-disciplinary literature data covering sociology, economics, political science, philosophy, education, law, and other disciplines has enabled them to establish a vast knowledge system, making it possible for them to propose policy recommendations that are more globally oriented and scientifically based[8]. Furthermore, with the aid of advanced natural language generation technology, LLMs can transform complex policy recommendations into clear and concise expressions, significantly enhancing the understandability and operability of the policy recommendations[9].

However, LLMs are currently in the early stages of development and still exhibit significant limitations, especially in basic aspects such as fact collection and verification. These tasks cannot yet be fully performed independently. This leads to the fact that no research has deeply discussed yet: if LLMs can fully master the relevant facts of specific social issues, can they propose corresponding policy recommendations, and what is the quality of these recommendations; to what extent LLMs can assist or even replace humans in completing policy suggestion work, and whether it is possible to reshape the decision-making thinking of human researchers and policy makers. These issues not only concern the application boundaries of LLMs, but also redefine epistemic foundations of social science inquiry.

In this context, we take the social security domain as an example to explore the capabilities of LLMs in generating policy recommendations. We focus on social security policy due to the following reasons. With the intensification of population aging, the development of diverse new business forms, changes in the pattern of wealth creation and distribution, and the continuous advancement of Chinese-style modernization, the improvement of China's social security policy system is facing unprecedented challenges[10]. The refinement of social security policies is closely linked to extensive public welfare and long-term national development. The scientific validity and rationality of policy recommendations in this area exert far-reaching influence on both social stability and governance efficiency. And technically, the data in the social security domain conforms to the data characteristics of LLMs that this study intends to investigate. Firstly, social security research focuses on real-world social problems as its research objects, and the domain has abundant policy practice data and field materials, which provide a textual basis for LLMs to conduct in-depth analysis and generate policy recommendations. Social security research exhibits inherent transdisciplinary complexity, spanning generational, institutional, and sectoral boundaries while addressing unstructured sociopolitical dynamics. Taking this domain as an example can reflect the ability characteristics of LLMs in making complex decisions when dealing with non-structural data. Specifically, this research aims to answer the following questions: (1) Can LLMs accurately understand the complex problems in social security domain and generate reasonable policy recommendations? (2) What are the similarities and differences in the policy recommendations generated by LLMs and those proposed by human researchers? (3) Do LLMs developed in different countries produce significant variation in policy recommendation outputs for identical social security issues? To address these questions, we conducted a comparative experiment to systematically evaluate the capabilities and limitations of LLMs in generating social security policy recommendations.

## 2 Related research and application exploration

### 2.1 A concise overview of LLM development

The emergence of large language models (LLMs) marks a transformative phase in artificial intelligence. Although AI technology has evolved over six decades, its applications remained largely confined to specialized domains such as computer vision, speech recognition, game-playing systems, and autonomous vehicles. LLMs represent a paradigm shift by demonstrating human-like knowledge integration and cross-domain reasoning capabilities for the first time, significantly expanding AI's societal applicability. In March 2023, OpenAI's GPT-4 breakthrough captured global attention. The model achieved unprecedented natural language fluency, enabling seamless human-machine communication. Moreover, in dealing with complex issues, GPT-4 has demonstrated a level close to that of human experts[11]. Subsequent 2024 releases (GPT-4o and GPT-4-turbo) further advanced task performance[12].

China's LLM development has progressed remarkably. Between 2024 and 2025, several institutions launched distinctive models: Moonshot AI's Kimi series excelled in long-context processing[13]; Baichuan Inc's models showed clinical decision-support advantages[14]; and DeepSeek's R1 model achieved internationally competitive reasoning performance[15]. These locally developed models exhibit optimized performance for Chinese language processing through specialized cultural and linguistic adaptation.

The cognitive capabilities of large language models (LLMs) primarily emerge during pretraining on vast corpora of unlabeled text data. Through this process, the models develop sophisticated representations of linguistic patterns, syntactic relationships, and semantic frameworks. To illustrate this scale, consider GPT-4's training corpus of approximately 13 trillion tokens[16]. This unprecedented volume in AI research equates to roughly 130 million standard-length books (assuming 100,000 tokens per book) - a dataset that dwarfs both conventional model training sets and the entire print collection of the U.S. Library of Congress (approximately 50 million cataloged items). The diversity of training data contributes significantly alongside data volume. Training on various content types including multiple languages, news articles, published works, research papers, technical documentation and online discussions helps language models develop cross-domain understanding. This varied training appears to support their ability to handle different kinds of tasks.

## 2.2 Research on LLMs in the social sciences

Current academic research on the intersection of large language models (LLMs) and social sciences reveals three primary research trajectories. The first examines the sociological framework for technology governance, particularly focusing on risk mitigation strategies during AI's integration into society[17-19]. This perspective advocates for a balanced regulatory approach that maintains technological development opportunities while implementing safeguards. Scholars have observed that human users often demonstrate uncritical reliance on AI systems in policy analysis and crisis management contexts. This finding underscores the need for dual governance mechanisms combining technical ethical adjustments with institutional innovations. Representative cases include the value alignment provisions in the EU AI Act and China's algorithmic filing system. These studies primarily emphasize applying social science principles to guide AI development, while largely overlooking how advances in AI technology might contribute to social science methodologies.

The second trajectory investigates the transformative potential of LLMs in reshaping disciplinary paradigms. Researchers are systematically examining how LLM capabilities can reconfigure established frameworks in public administration, economics, and communication studies[20-25]. In public administration, for instance, LLMs demonstrate capacity for evidence-based policy support through data synthesis, service delivery optimization via automated systems, governance model innovation through predictive analytics. However, LLMs may also lead to issues such as decision-making failure due to technology abuse, social inequality caused by technology monopolies, and security crises due to data exploitation. Nevertheless, most of these studies remain at the theoretical deduction stage and lack empirical research support.

The third trajectory focuses on LLMs application evaluation, assessing model performance across various real-world tasks. Deroy *et al*.[26] tested if LLMs can automatically generate abstractive summaries for Indian court case judgements. Nay *et al*.[27] tested LLM capabilities in retrieving and utilizing the relevant legal authority. Ziems *et al*.[28] measured the performance of 13 LLMs on summarizing relevant aspects, elucidating the hidden social meaning behind a text and implementing social theory by restructuring an utterance in computational social science. Yu and Wang [29] tested GPT-3.5 on news verification, finding superior accuracy for political/authentic news but deficiencies with non-political/fake news, plus generational and linguistic biases. These studies assessed LLM performance in narrow, structured tasks without addressing the generation competency of LLMs for complex unstructured tasks in policy refinement scenarios.

Overall, existing studies have demonstrated innovations in two major directions: treating LLMs as research objects (social science of AI) and utilizing LLMs as research tools (AI for social science). These contributions have constructed novel theoretical frameworks for AI algorithmic cognition and established methodological foundations for interdisciplinary research. However, these studies also exhibit notable limitations. There remains a lack of empirical assessment regarding LLMs' capabilities in systematic diagnostics, value trade-offs, risk/cost prediction, which are precisely the core competencies of social scientist. Therefore, this research focuses on evaluating LLMs' capacity to generate policy recommendations based on existing research materials of social issues. Through comparative testing and analysis, it aims to provide practical frameworks for human-AI collaboration in social science research.

## 3 Methodology

### 3.1 Evaluation Approach for LLMs Policy Recommendation Capabilities

Building on existing LLM evaluation methodologies, this study adapts assessment frameworks from clinical medicine. The adopted approach mirrors established medical diagnostic protocols where both human physicians and LLMs respond to identical open-ended case studies, with subsequent qualitative evaluation by expert panels using non-standardized criteria[2,30,31]. This comparative method enables holistic identification of model capabilities and limitations. For example, Stanford University and Harvard University jointly conducted an open-ended test with six clinical cases, and experts freely evaluated the strengths and weaknesses of the model. The expert assessment revealed that while LLMs demonstrated superior comprehensiveness in identifying rare diseases, they exhibited notable limitations including verbose questioning patterns and inconsistent output reliability.

This study employs a comparable evaluation methodology, utilizing a context-embedded generation-adaptation approach to assess LLMs' policy recommendation capabilities in social security. The research design involves selecting multiple academic papers and redacting their original policy recommendations while preserving all contextual information, thereby providing LLMs with comprehensive background and situational details about the relevant social issues. The models are then required to generate new policy recommendations based on this contextual information. Expert evaluators subsequently compare these AI-generated recommendations with the original human-authored proposals, assessing their overall adaptability and identifying strengths and limitations in the LLMs' outputs.

Recognizing that social security research narratives are deeply influenced by historical, cultural, and ethical factors, the study incorporates two leading LLMs for comparative analysis: OpenAI's GPT-4o (representing the U.S. model) and DeepSeek-R1 (representing China's domestic model). This comparative framework serves dual purposes: first, it evaluates the technical proficiency of current state-of-the-art language models; second, it examines potential variations in cultural adaptability - specifically, the models' capacity to comprehend nuanced aspects of policy ethics, cultural traditions, resource allocation patterns, and related contextual factors within the same application scenario. The methodology not only benchmarks model performance but also provides insights into how different cultural and developmental contexts may influence AI-generated policy recommendations.

The adoption of medical evaluation methodologies in this study is justified by several key considerations. First, there exists a fundamental structural homology between policy recommendation scenarios in social sciences and medical diagnostic processes, particularly in terms of data characteristics, cognitive frameworks, and decision-making paradigms. Both domains deal with unstructured data derived from complex systems - whether patient case histories in healthcare or multifaceted social problem descriptions in policy analysis. These scenarios demand comprehensive assessments that simultaneously consider professional accuracy, practical implementation feasibility, and ethical alignment. Moreover, both fields follow a similar progressive reasoning logic moving from problem representation to evidence analysis and finally solution generation, making methodological transfer between the domains conceptually sound. Second, medical research has established pioneering and authoritative frameworks for LLM evaluation that have gained widespread acceptance in artificial intelligence research. Leading institutions have implemented this approach with their findings published in top-tier medical journals such as Nature Medicine and JAMA. The validity assessment approach further supports this methodological choice - medical evaluations employ open-ended assessment protocols rather than predefined evaluation metrics, utilizing a "free response-feature induction" approach. Compared to standardized evaluation frameworks, this method better avoids biases introduced by incomplete or restrictive predefined indicators while more effectively capturing emergent LLM capabilities and unanticipated limitations in policy recommendation tasks. The subsequent section will elaborate on the specific evaluation methodology and implementation steps employed in this research.

### 3.2 Dataset selection

This research selects the papers from the *Chinese Social Security Review* as the dataset for evaluation. On one hand, this journal holds significant academic authority and policy influence. As a CSSCI-indexed journal and the flagship publication of the China Social Security Association, its articles feature in-depth examinations of China's social realities and social security reforms, providing critical practical guidance. Using its research papers as comparison benchmarks ensures LLMs evaluations adhere to high-quality standards. Moreover, the journal maintains thematic focus, covering core social security issues (pensions, healthcare, fertility, employment) and emerging topics (social security adaptation in the digital economy) across over a dozen policy scenarios. Evaluating LLMs against these studies helps avoid systematic bias caused by domain randomness.

The specific process of selecting papers was as follows:

   a) Candidate pool selection: All 61 papers published in *Chinese Social Security Review* in 2024 were initially selected as candidates.
   b) Preliminary screening: Papers containing purely theoretical analyses or lacking new empirical findings were excluded, leaving 37 papers.
   c) Secondary screening: Articles with minimal or overly generalized policy recommendations were removed, resulting in 28 remaining papers.
   d) Final selection: 10 papers were randomly selected from the remaining pool while ensuring maximum thematic coverage.

The sample size determination was primarily constrained by the organizational costs of expert evaluation - each assessment required multiple experts to thoroughly review and compare materials while providing detailed comments, with research costs increasing exponentially with additional cases. Following the precedent of JAMA (a leading medical journal) which utilized 6 cases in its LLM evaluation [35], this study ultimately established a core test set comprising 10 papers (see Table 1 for details). It should be noted that 10 samples are insufficient to establish a comprehensive evaluation system, but adequately serve this exploratory study's fundamental purpose: to reveal basic characteristics and underlying patterns of LLMs in policy recommendation generation through empirical cases.

### 3.3 Generating policy recommendations

The specific steps for generating policy recommendations by the selected LLMs were as follows. All experimental interactions were originally conducted in Chinese.

- a) Data Preparation: The chapters preceding the policy recommendations were extracted from target papers as input data. For instance, in the paper "Analysis of Utilization Status and Optimization Strategies for Home-Based Community Elderly Care Facilities in China" (published in *Chinese Social Security Review*, Issue 1, 2024), which consisted of five chapters: (1) Introduction, (2) Conceptual Framework, (3) Descriptive Analysis, (4) Regression Analysis, and (5) Conclusions and Recommendations, the first four chapters were selected as input.

- b) Prompt Design: A model prompt was designed to clearly define the task:
  *"Read the provided research excerpt, conduct a comprehensive analysis of the social issues and empirical findings presented, and propose actionable policy recommendations."*

- c) Input Instruction and Output Recording: The prompt was concatenated with the extracted paper content to form complete input text. This text was then fed into the LLM. To ensure the model relied solely on its intrinsic capabilities, internet search functionality was disabled during inference. The model's outputs were systematically recorded.

- d) Experimental Procedure: The above steps were repeated for all 10 papers in the test set using both models (GPT-4o and DeepSeek-R1) until complete results were obtained. Detailed outputs are presented in Appendix.

Through this process, we can systematically evaluate the performance of LLMs in the task of policy recommendation generation and ensure the consistency and reproducibility of the experiment.

### 3.4 Comparing the experimental results

This research refers to the evaluation research on the medical diagnosis advice capabilities of LLMs mentioned in the foregoing text and adopts the blind review method, inviting domain experts to evaluate the experimental results. The evaluation work mainly focuses on two core objectives: (1) Making qualitative judgments on the quality of policy recommendations; (2) Systematically summarizing the characteristic performances of LLMs in policy recommendation tasks, including their advantages and limitations.

To ensure the objectivity of the evaluation, this research adopts an open evaluation framework and avoids pre-setting evaluation dimensions (such as the comprehensiveness and feasibility of policy recommendations), to prevent the generation of result deviations caused by evaluation standards. Specifically, experts are asked to directly rank the policy suggestions generated by human researchers and multiple LLMs instead of scoring each individual model's output results. This design can effectively avoid the risk of amplifying or masking model characteristics due to preset dimensions. Experts need to comment on the advantages and disadvantages of each policy recommendation. Based on this, researchers conduct a systematic analysis of all comments and extract representative evaluation labels, and compare the performance differences between LLMs and human researchers in the policy suggestion task through frequency statistics.

The detailed evaluation procedure was as follows:

- a) Three domain experts were selected for participation;

b) Experts reviewed paper excerpts (with recommendations redacted) to understand research context;

c) Experts evaluated three randomly ordered policy recommendations (containing both human-authored and model-generated versions);

d) Experts ranked the recommendations, allowing for ties (e.g., 1st, 1st, 3rd or 1st, 2nd, 2nd);

e) Experts provided written critiques detailing strengths and weaknesses for each recommendation

f) Steps b-e were repeated for all 10 papers;

g) Three experts independently completed the entire evaluation process;

h) The research team consolidated all expert critiques to extract representative evaluation labels;

i) Frequency analysis was conducted on the identified labels to quantify performance differences between LLMs and human researchers.

## 4 Analysis of evaluation results

### 4.1 The ranking of model performance

Before releasing the comparison results, it is necessary to further elaborate on the details of the evaluation setup to ensure process rigor and result fairness. First, during the testing phase, the author noted that AI-generated policy recommendations are typically presented as bullet-point lists, whereas human researchers' policy recommendations often take the form of detailed paragraph explanations. This discrepancy could enable evaluation experts to easily distinguish between AI and human outputs. To eliminate this bias, we standardized the format of the human researchers' recommendations by condensing their paragraph explanations into bullet-point form to match the AI output format. Specifically, the formatting adjustment involved two steps: the first step was to conduct preliminary format conversion using DeepSeek-R1; the second step was for independent human experts to compare pre- and post-conversion content to ensure semantic consistency and correcting inconsistencies. Notably, the experts involved in formatting verification were a distinct group from those conducting the final evaluation to avoid potential bias.

Second, to ensure evaluation objectivity, we randomly reordered the three sets of recommendations for each paper (including outputs from two models and human researchers), so that the results from human researchers, GPT-4o, and DeepSeek-R1 appeared in different positions each time. This measure effectively avoided subjective bias from evaluation experts due to the order of recommendations.

Finally, during the statistical phase, we employed a rank-order scoring method for evaluation. Specifically, each expert independently ranked the three sets of recommendations for each paper, translating these rankings into scores (i.e., 1 point for first place, 2 points for second place, and 3 points for third place). We then calculated the average score for each candidate and re-ranked them (from lowest to highest) based on these averages to determine the final evaluation results. This approach not only ensured fairness in the assessment but also enhanced the statistical robustness of the outcomes.

The final ranking statistics results are shown in **Table 1**. The two models showed significant differences in their rankings: China's domestically developed DeepSeek-R1 model achieved a notably leading position, with 9 out of 10 policy recommendations ranking first, surpassing both human researchers and GPT-4o. In contrast, as one of the world's leading model series, GPT-4o performed significantly behind human researchers and DeepSeek R1 in this evaluation, with 9 out of 10 recommendations ranking lower than both, and only one recommendation securing second place. Human researchers consistently ranked between DeepSeek-R1 and GPT-4o. Based on these results, the authors selected one case where DeepSeek-R1 outperformed human researchers and one case where GPT-4o underperformed human researchers for detailed analysis, to demonstrate the respective characteristics of the two models compared to human researchers.

**Table 1 Ranking of policy recommendations by Human researchers, GPT-4o, and DeepSeek-R1**

| Paper Sample | Human researchers | LLMs | |
| --- | --- | --- | --- |
| | | GPT-4o | DS-R1 |
| Fertility Support Policy under the "Dual Goal": Construction Logic and Implementation Path[32] | 2 | 3 | 1 |
| Reflections on the Multi-Pillar Pension Approach[33] | 3 | 2 | 1 |
| Efficiency Measurement, Spatial Network Structure Characteristics, and Influencing Factors of Elderly Care Service Institutions in the Context of High-Quality Development[34] | 2 | 3 | 1 |
| Family Support Patterns for Disabled Elderly and Policy Implications [35] | 1 | 3 | 2 |
| Utilization and Optimization of Home and Community-Based Elderly Care Facilities in China[36] | 2 | 3 | 1 |
| "Market Contract System" in the Supply of Rural Elderly Services: Operation Model and Practical Logic[37] | 2 | 3 | 1 |
| Social Insurance Participation Choices of Workers in New Business Forms and Their Influencing Factors: A Survey of Delivery Riders and Couriers[38] | 2 | 3 | 1 |
| The Legal Dilemma and Reform Reconstruction of the Individual Account in Basic Medical Insurance for Employees[39] | 2 | 3 | 1 |
| 25 Years of Reform in China's Medical Insurance Payment System: Achievements, Problems, and Prospects[40] | 2 | 3 | 1 |
| Navigating Social Security Challenges in Flexible Employment[41] | 2 | 3 | 1 |
| Total (average ranking) | 2 | 3 | 1 |

**Table 2** presents a representative case from "Social Insurance Participation Choices of Workers in New Business Forms and Their Influencing Factors: A Survey of Delivery Riders and Couriers" where the LLM outperformed human researchers in a specific policy scenario. The comparative analysis reveals that both the AI system and human experts demonstrated professional competence in proposing comprehensive social security reforms, including legislative improvements, regulatory enhancements, and insurance scheme optimizations. However, notable differences emerged in the level of operational specificity. The DeepSeek-R1 model exhibited superior capacity for actionable policy recommendations compared to human researchers. For instance, while human proposals focused on conceptual frameworks like tiered participation systems and single-category insurance models, the AI generated detailed implementation plans with quantified parameters and international benchmarks. A concrete example appears in low-income protection measures, where DeepSeek-R1 specified that workers earning below 60% of local average wages could have their pension contribution base set at 40-50% of standard rates, while also referencing Germany's Mini-job system as a potential model for exempting low-income groups from social security contributions. This contrast illustrates the complementary value of AI systems in translating policy principles into executable measures.

Regarding regulatory strengthening, DeepSeek-R1 proposed a relatively comprehensive oversight system encompassing regulatory platform development, technological tool implementation, and clear

**Table 2 A typical case in which LLM outperform human researchers in policy recommendations within social security scenarios**

|  | Human researchers | DeepSeek-R1 |
|---|---|---|
| Improving Laws and Regulations | **Strengthen Identification of Labor Relations**<br>- Strictly define labor relations based on employment facts, and implement categorized protection measures. | **Establish a "Three-part Classification of Labor" Legislative System**<br>-Revise the Labor Law or formulate the Law on the Protection of the Rights and Interests of Workers in New Employment Forms, clarify the legal status of "incomplete labor relations", and grant them social security enrollment rights.<br>-**Reference case:** The EU "Digital Labor Platform Directive" (2024) regards platform workers as "quasi-employees", who enjoy partial social security rights. |
| Strengthening Supervision | **Strengthen Labor Supervision**<br>- Focus on cracking down on platform enterprises' illegal subcontracting and acts of disregarding labor relations facts.<br>- Implement flexible supervision within the legal framework for situations that do not fully meet labor relations.<br>- Promote local practical experience:<br>  Entry supervision: Require platforms to sign contracting agreements with high-quality contractors.<br>  Process supervision: Establish a regular verification mechanism, and regularly check the implementation of labor contracts. | **Construct a "Platform agency payment + Government Supervision" System**<br>- Compel platforms to connect to the national social security public service platform, automatically deduct and pay social security fees (requiring authorization from practitioners), with human resources and social security departments monitoring payment data in real time.<br>- **Technical tools:** Use blockchain technology to record data such as order volume and working hours, as auxiliary basis for determining insurance eligibility.<br>**Supervision reinforcement: Implement enterprise main responsibility**<br>- **Carry out "penetrative employment supervision"**<br>Require platform enterprises to publicly list cooperative contractors, prohibit transferring employment responsibilities through subcontracting at four levels or above. Impose fines of 2-3 times the payable social security fees for acts of evading labor relations.<br>- **Establish a "Platform Enterprise Credit Rating" system**<br>Incorporate the compliance of social security payments, workers' complaint rates, etc., into enterprise credit evaluation. Low ratings restrict participation in government procurement and financing support. |
| Insurance System | **Full Coverage of Occupational Injury Protection**<br>- Break through the mandatory binding of work-related injury insurance and employers.<br>- Promote pilot experiences:<br>  Government-subsidized commercial insurance<br>  Priority coverage of work-related injury insurance for grassroots express delivery outlets<br>  Single-insurance-type participation model for specific groups. | **Implement the "Single-insurance-type Participation in Social Insurance" Model**<br>- Allow practitioners in new business forms to participate in insurance by type as needed (such as separately participating in work-related injury insurance), reducing enterprise cost pressure.<br>**Upgrade local pilots:** Promote the single-insurance-type participation policy for work-related injury insurance in Zhejiang and Guangdong to the whole country, covering all platform enterprises. |
| Insurance Participation Mechanism | **Establish a tiered insurance participation mechanism:**<br>- Compel low-income earners to participate in resident insurance<br>- Guide middle and high-income earners to participate in employee insurance<br>Implement flexible payment policies that match the income characteristics of new business forms. | **Establish a "Platform-Government-Individual" Social Security Expense Sharing Mechanism**<br>- Platforms bear certain social security fees according to order volume or income proportion (such as 30% of work-related injury insurance fees). The government provides tax incentives, and individuals pay the remaining part at the flexible employment premium rate.<br>**Differentiated payment design**<br>- Implement tiered premium rates for low-income practitioners in new business forms: If the monthly income is lower than 60% of the local average social wage, the payment base for employee pension insurance can be verified at 40%-50%.<br>- **Learn from experience:** Germany's Mini-job system allows those with a monthly income of less than 520 euros to be exempt from social security payments, with employers bearing fixed expenses. |
| Value Incentives | Not mentioned | **Strengthen Policy Incentives**<br>- For flexible employment personnel with continuous insurance participation for 10 years, increase the pension calculation ratio by 5%; provide annual tax rebate rewards (such as 5% of the payment amount) for those who pay in full and on time.<br>**Implement the "Social Security Rights and Interests Points System"**<br>- Link social security payment records with urban public services (such as children's enrollment, public rental housing applications). Those who meet the points standard can enjoy priority rights and interests.<br>**Strengthen Hedging against Adverse Selection Risks**<br>- Introduce the "Young and Healthy Incentive Fund" in employee medical insurance. Refund 20% of the annual premium to those aged 25-35 who have not claimed reimbursement for three consecutive years, balancing the risk pool structure. |
| Social Security Transfer | Open cross-regional and cross-system social security transfer and continuation channels. | **Open Urban-Rural Social Security Transfer and Continuation Channels**<br>- Establish a national social security transfer platform, allowing the payment years of urban and rural resident pension insurance and employee pension insurance to be converted proportionally.<br>- Achieve provincial-level overall planning for the cross-provincial and off-site medical reimbursement catalog, and gradually transition to a national unified reimbursement standard. |
| General Systems | Extend the minimum payment years for pension insurance Improve the outpatient co-ordination mechanism for medical insurance. | Not mentioned |
| Convenience Services | Not mentioned | **Carry out the Construction of "Social Security Service Stations"**<br>- Set up self-service terminals at express delivery sites and food delivery stations, providing one-stop services such as insurance participation, inquiries, and benefit claims. |
| Implementation Path | Not mentioned | - Short-term (1-2 years): Expand the pilot of occupational injury protection, and complete the solicitation of opinions on the revised draft of the Labor Law.<br>- Medium-term (3-5 years): Complete the national social security transfer and continuation system, and fully implement the single-insurance-type participation model.<br>- Long-term (more than 5 years): Achieve a complete decoupling of social security and labor relations, and establish a national social security system based on occupational risks. |
| Feasibility Assessment | Not mentioned | - Controllable fiscal cost: Tiered premium rates and tax incentives can be offset by expanding the insurance participation base.<br>- Mature technical support: The national social security information platform has achieved provincial-level overall planning. Blockchain and big data technologies can assist in policy implementation.<br>- Less social resistance: Policy design considers both enterprise costs and workers' rights and interests. Platforms can digest part of the costs through economies of scale. |

accountability mechanisms. The model specified actionable provisions such as "imposing fines equivalent to 200-300% of owed social security contributions for labor relationship circumvention." Regarding policy innovation, DeepSeek-R1 introduced two novel dimensions: value incentive mechanisms and user-friendly service systems. The incentive framework combines monetary and service rewards, proposing measures like a 5% pension benefit increase for flexible workers maintaining 10 years of continuous coverage, annual tax rebates equivalent to 5% of contributions for timely full payments, and linking social security records with access to urban public services (e.g., school admissions, public housing applications). Additionally, it suggested installing self-service terminals at delivery hubs to provide gig workers with integrated services including enrollment, payment tracking, and benefit claims. These measures demonstrate potential to enhance participation rates while mitigating fiscal pressures from incentive programs. The model further distinguished itself through detailed implementation roadmaps featuring phased short-, medium-, and long-term objectives with clear timelines, an aspect notably absent from human researchers' proposals. This systematic approach enhances both the practicality and predictability of the policy recommendations.

**Table 3** presents a typical case where policy recommendations from the GPT-4o model fell short compared to those proposed by human researchers. This case is drawn from "Family Support Patterns for Disabled Elderly and Policy Implications". In response to the caregiving dilemmas of disabled elderly revealed in this paper, human researchers did not limit themselves to "symptomatic treatment" solutions. Instead, they constructed a comprehensive policy framework based on conceptual transformations. Its core philosophy involves three shifts: First, emphasizing the family as the policy-making unit rather than individual seniors, to holistically address the dual needs of disabled elders and their caregivers; second, implementing phased intervention strategies including early professional guidance, mid-term multidimensional support, and late-stage institutional referrals; third, establishing preemptive intervention mechanisms to prevent family disruptions caused by sudden disability. For example, in family policy areas, human researchers proposed establishing paid care leave, visiting leave for elderly relatives in other localities, and a system for agency payment of long-term care insurance for family caregivers. In contrast, GPT-4o's policy recommendations demonstrated an obvious "symptom-focused" approach. In terms of conceptual transformation, it merely mentioned strengthening propaganda and attention to filial piety culture and disability awareness. Such "superficial appeals" neither address the deep cultural contradictions in intergenerational relationship restructuring during the individualization process, nor resolve institutional-level responsibility allocation issues. While GPT-4o incorporates Chinese traditional cultural ethics into its policy recommendation framework, it fails to translate these into actionable institutional designs. For another example, in enhancing economic support, GPT-4o only proposed lowering subsidy eligibility thresholds for disabled seniors and increasing pension insurance coverage, without mentioning considerations for other family members, lacking the holistic perspective of family policies suited to China's socio-political context.

In comparison with human researchers, GPT-4o's policy recommendations demonstrate a notable lack of precise targeting and concrete implementation measures. Taking the improvement of social support systems as an example, human researchers provided detailed specifications for each policy component: regarding facility optimization, they explicitly proposed adjusting the ratio between specialized care facilities and standard institutions, enhancing smart technology integration in institutional care, and supporting accessibility modifications for disabled elders' homes. Concerning human resources, they outlined practical incentive measures such as special post allowances for care professionals. In contrast, GPT-4o merely suggested generic directions like developing community care centers, dementia-specialized institutions, and family support services, along with increasing professional staffing and caregiver training—all without specifying operational details or implementation pathways. This contrast reveals GPT-4o's current limitations in translating broad policy concepts into actionable plans within local social governance contexts.

Moreover, GPT-4o's recommendations exhibit significant shortcomings in cost assessment and resource allocation considerations. For instance, its proposal to expand disability benefits for elderly populations fails to address funding sources or fiscal sustainability constraints, rendering the suggestion largely impractical. Similarly, while advocating for expanded community care services through increased facilities and professional staffing, the model demonstrates limited understanding of China's structural workforce challenges in the care sector - specifically the existing 3-million shortage of qualified personnel and their uneven geographical distribution. This reveals GPT-4o's current limitations in incorporating regional-specific operational constraints into policy formulation, particularly when dealing with localized resource allocation issues and implementation feasibility. The recommendations consequently lack grounding in the practical realities of human resource availability and spatial distribution patterns within the care industry.

**Table 3 A typical case in which LLMs perform less well than human researchers in offering policy recommendations in social security scenarios**

| | Human Researchers | GPT4o |
|---|---|---|
| Policy Core Concept Transformation | - Formulate policies based on families, considering the needs of both disabled elderly and caregivers<br>- Implement phased intervention strategies (early professional guidance, mid-term multidimensional support, late-stage institutional referral)<br>- Establish a pre-emptive intervention mechanism (risk prevention + capability building) | Not mentioned |
| Direct Economic Support | Not mentioned | **Expand the subsidy scope for disabled elderly**: Lower subsidy thresholds and increase economic support for families of disabled elderly. |
| Economic Security System | **- Long-term care security system**<br>- Achieve full coverage of long-term care insurance, unify payment standards by disability level<br>- Establish a lifelong care mechanism (covering multiple main bodies such as institutions/families/society)<br>- Establish a "fee follows the person" long-term care payment mechanism<br>  Improve economic support packages:<br>  Raise individual tax deduction standards for the elderly<br>  Set up caregiver subsidy systems<br>  Implement care insurance delegation systems<br>  Promote the complementary connection between preferential care and nursing subsidies | **Increase pension insurance coverage**: Especially for rural areas and low-income families, ensure disabled elderly have stable income sources.<br>**Pilot long-term care insurance**: Explore long-term care insurance systems suitable for China's national conditions to share family economic pressure.<br>**Optimize legal safeguards**: Improve relevant laws and regulations, clarify the responsibility allocation of the government, society, and families in elderly care. |
| Social Support System | Hierarchical service system<br>- Family care support system:<br>  Respite services (knowledge training / alternative care / psychological counseling / professional medical care) care/psychological counseling/professional medical care)<br>  Paid nursing leave system (classified by disability level)<br>  Construct a family resilience assessment and intervention system<br>- Special group protection:<br>  Empty nest/lonely/disabled elderly special services<br>  Establish service supervision and patrol mechanisms<br>  Explore non-family caregiver treatment payment | **Develop community care**: Increase community care stations and professional caregivers, provide door-to-door services.<br>**Establish specialized care institutions for disability**: Provide specialized services for disabled elderly, reduce family care burdens.<br>**Provide caregiver training**: Offer care skills and psychological support training through community and online platforms.<br>**Set up family support centers**: Provide information consultation, psychological counseling, and temporary care for disabled elderly families. |
| Relevant Supporting Measures | - Optimize facility structure:<br>  Increase the proportion of nursing-type beds in elderly care institutions, reduce the proportion of ordinary beds<br>- Talent team construction:<br>  Professional cultivation in colleges (tuition exemption + school-running subsidies)<br>  Vocational incentives (post subsidies + social honors)<br>- Smart technology application:<br>  Promote home smart monitoring systems<br>  Build regional health data platforms<br>  Develop aging-friendly smart aids | - Advocate filial piety culture: Combine modernization needs, encourage social support and respect for family elderly care.<br>- Raise social awareness: Through media and community activities, enhance public awareness of disability. |

### 4.2 Tag-based statistical analysis of model characteristics

After evaluating the overall baseline level of LLMs, this research further extracted labels from expert comments to examine the commonalities and characteristics in policy recommendations between human researchers and LLMs (GPT-4o and DeepSeek-R1). By selecting the 8 most frequently occurring labels from the advantage and disadvantage labels (see **Table 4**), a radar chart was created to visually display the characteristic performance of the two models in policy recommendations.

As shown in the radar chart (see **Figure 1**), overall, human researchers exhibit relatively balanced strengths, with only a slight underperformance in the aspect of clear quantitative objectives. In contrast, the DeepSeek-R1 and GPT-4o models demonstrate distinct areas of advantage, excelling particularly in their specialized domains. Specifically, DeepSeek-R1 outperforms human researchers in the systematic design of institutions (multi-dimensional considerations, multi-department collaboration), institutional innovation, concrete measures, and quantitative objectives. It approaches the level of human researchers in terms of applicability (coverage of target groups), but falls short in operability, short-term effectiveness, and cost and risk control. Comparatively, GPT-4o shows a slight advantage in short-term effectiveness, but its performance in other areas is not particularly outstanding.

Regarding limitations, DeepSeek-R1's proposals typically face greater implementation barriers, higher costs and elevated risks compared to human researchers' recommendations. In contrast, GPT-4o's suggestions demonstrate weaker policy impact (characterized by limited policy instruments), vague content, and operational challenges at grassroots levels. This comparative analysis reveals distinct but equally significant constraints in both AI systems' policy formulation capabilities when benchmarked against human expertise.

**Table 4 Description of high-frequency advantage labels and disadvantage labels**

| Advantage Labels | Explanation | Disadvantage Labels | Explanation |
| --- | --- | --- | --- |
| Systematic and Comprehensive, Consideration of Multiple Dimensions, Coordination Among Multiple Departments | The policy covers multiple aspects horizontally or involves multiple levels vertically, considers coordination among multiple departments, addresses root problems, and has good long-term effects. | High Coordination Difficulty, Great Implementation Resistance | Policy implementation requires coordinating multiple departments, with high coordination pressure; there is significant policy inertia and great reform resistance. |
| Specific Measures | There are specific measures or examples of measures. | High Cost, Great Fiscal Pressure | High resource demand and cost, lacking consideration for financial resources. |
| Strong Operability, With Case References | The measures are highly feasible, with details, steps, and reference cases. | Slow Effect, Long Cycle | Policy effects are not easily reflected, hard to assess, or require a long time. |
| Wide Coverage of Groups, Large Applicable Scope | The policy adopts hierarchical design, considering different groups. | High Difficulty in Grassroots Implementation | High requirements for grassroots execution capabilities; lacking incentives. |
| Balancing Costs, Balancing Risks | There are considerations for financial resources and risk control measures. | General/Vague/Slogan-like | Lacks detailed policy means or specific measures. |
| Short-Term Effect | Direct effect, fast in showing results. | Small Policy Strength, Single Policy Tool | Treats symptoms but not the root cause, overly reliant on laws and regulations or central fiscal appropriations. |
| Clear Quantitative Objectives | There are clear quantitative suggestions. | Narrow Applicable Scope, One-Size-Fits-All Approach | Single perspective, narrow applicable population. |
| Innovative | The policy or measures are novel. | Inadequate Risk Control, Many Remaining Issues | Lacks risk considerations and necessary risk control measures. |

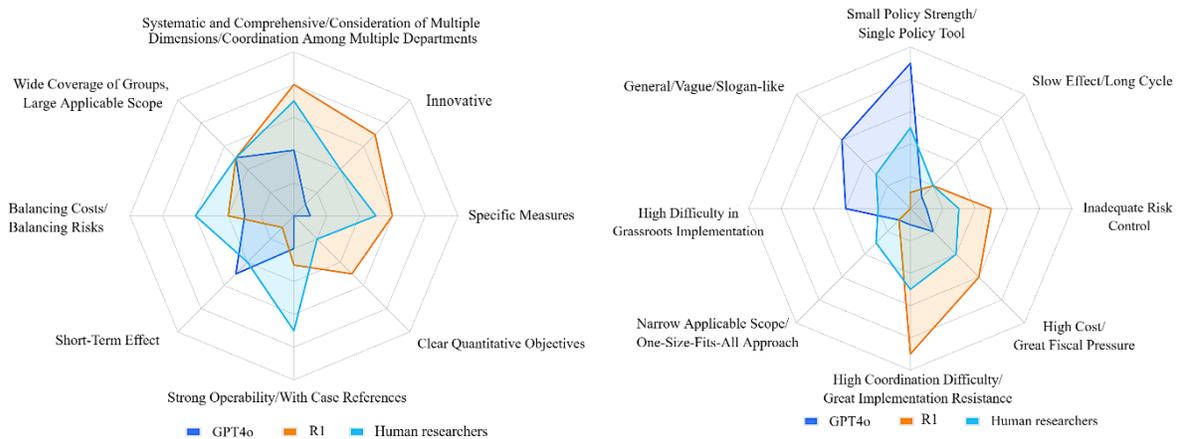

**Figure 1 The characteristic distribution of policy recommendations offered by human researchers, GPT-4o, and DeepSeek-R1**

## 5  Conclusion and Discussion

This study compares policy recommendations on social security proposed by human researchers and large language models (LLMs) - DeepSeek-R1 and GPT-4o. The findings reveal that DeepSeek-R1 demonstrates superior innovative capacity compared to human researchers, generating complex, systemic reform proposals featuring multi-departmental coordination, quantified targets, and phased implementation roadmaps. However, it underperforms human researchers in cost control, resistance anticipation, and risk management - a paradoxical capability profile reflecting fundamental differences between artificial and human intelligence: the former relies on pattern extrapolation from massive data, while the latter draws on practical wisdom grounded in social realities. Furthermore, the domestic model DeepSeek-R1 outperformed the U.S.-developed GPT-4o in this experiment, producing more comprehensive and feasible recommendations that better incorporate local socioeconomic contexts. This performance gap reveals differences in the models' capacity to recognize deep semantic features related to policy ethics, cultural traditions, and resource endowments, suggesting that region-specific training data influences policy recommendation appropriateness.

These characteristics reflect an inherent tension between LLMs' technical properties and social security research paradigms. The models' systemic reform design capability stems from two technical features: First, their training on trillions of cross-disciplinary tokens enables extracting and integrating knowledge from fragmented information, allowing them to transcend human experts' domain limitations and identify implicit connections across healthcare, finance, and civil affairs systems. Second, lacking embodied cognition in real-world contexts, their policy generation process disregards practical constraints like interest conflicts and implementation barriers. This decontextualized characteristic forms the technical basis for systemic innovation, enabling idealized institutional designs under "frictionless assumptions." However, this same feature leads to structural deficiencies in feasibility as the models prioritize textual coherence over practical implementation.

The study has several methodological limitations. Due to resource constraints, the sample only covered ten papers, leaving some social security policy scenarios unexamined. The absence of interdisciplinary expert panel discussions may introduce subjective bias in evaluations. Additionally, the qualitative assessment approach cannot quantify performance differences between LLMs and human researchers across core policy recommendation competencies. Nevertheless, as preliminary exploration of LLMs' policy advisory capabilities in social science research, this study has fulfilled its exploratory mission by delineating LLMs' capabilities relative to human researchers, leaving remaining questions for future research.

Returning to our initial question: Can LLMs become new partners in policy refinement? The results suggest LLMs possess unique advantages in policy research through global knowledge integration that overcomes human cognitive path dependence, revealing hidden connections beyond disciplinary boundaries. However, uncritical adoption of model-generated recommendations risks "technical rationality suspension" - theoretically sound solutions ungrounded in real social structures or institutional environments. By strategically leveraging complementary strengths through a three-phase collaborative model - human-led problem definition leveraging contextual awareness, AI-augmented solution generation for diverse options, and expert-driven evaluation incorporating professional judgment - LLMs could become valuable policy optimization assistants. Future development should explore integrating LLMs with other technologies (e.g., predictive modeling, simulation) while embedding dynamic constraint awareness (e.g., fiscal pressures, implementation resistance, resource disparities) to achieve dialectical unity between policy ideals and practical feasibility.

## References

[1] Wang Y, LI Q, DAI Z, XU Y. (2024). Current status and trends in large language modeling research. *Chinese Journal of Engineering*, 46(8): 1411-1425.

[2] Goh, E., Gallo, R., Hom, J., Strong, E., Weng, Y., Kerman, H., ... & Chen, J. H. (2024). Large language model influence on diagnostic reasoning: a randomized clinical trial. *JAMA Network Open*, 7(10), e2440969-e2440969.

[3] "AI Civil Servants Are Coming: How Will They Transform Grassroots Governance?"(2025-02-18)[2025-02-25]https://news.youth.cn/jsxw/202502/t20250218_15834100.htm

[4] Gielens, E., Sowula, J., & Leifeld, P. (2025). Goodbye human annotators? Content analysis of social policy debates using ChatGPT. *Journal of Social Policy*, 1-20..

[5] Chen Shuisheng (2020). What Constitutes 'Good Policy'? A Systematic Review of Public Policy Quality Research. *Journal of Public Administration*,13(03):172-192+200.（in Chinese）

[6] Ibrahim, N., Aboulela, S., Ibrahim, A., & Kashef, R. (2024). A survey on augmenting knowledge graphs (KGs) with large language models (LLMs): models, evaluation metrics, benchmarks, and challenges. *Discover Artificial Intelligence*, *4*(1), 76.

[7] Qiu, X., Sun, T., Xu, Y., Shao, Y., Dai, N., & Huang, X. (2020). Pre-trained models for natural language processing: A survey. *Science China technological sciences*, *63*(10), 1872-1897.

[8] Jiang L, GU J. (2024). Research on the Interdisciplinary Knowledge Organization and Visualization of Literatures Under the Background of Large Language Model. *Library and Information Service*, (23):17-29. (in Chinese)

[9] Chaleshtori, F. H., Ghosal, A., Gill, A., Bambroo, P., & Marasović, A. (2024). On Evaluating Explanation Utility for Human-AI Decision Making in NLP. *arXiv preprint arXiv:2407.03545.*.

[10] Zheng Gongcheng. (2024). The Outline of the Social Security System with Chinese Characteristics. *Chinese Social Security Review*, 8(01):3-22.( in Chinese)

[11] OpenAI, "GPT-4 Technical Report", arXiv preprint, 2024, arXiv:2303.08774.

[12] OpenAI, "GPT-o1 System Card", arXiv preprint, 2024, arXiv:2412.16720.

[13] Kimi Team. (2025). Kimi k1.5: Scaling Reinforcement Learning with LLMs. *arXiv preprint,* arXiv:2501.12599.

[14] Baichuan Team. (2025). Baichuan-Omni-1.5 Technical Report. *arXiv preprint,* arXiv:2501.15368.

[15] DeepSeek-AI. (2025). DeepSeek-R1: Incentivizing Reasoning Capability in LLMs via Reinforcement Learning. *arXiv preprint*, arXiv:2501.12948.

[16] Restack. (2025). Gpt4 Training Dataset Size Insights, https://www.restack.io/p/gpt-4-training-answer-dataset-size-cat-ai

[17] Karell, D., Sachs, J., & Barrett, R. (2025). Synthetic duality: A framework for analyzing generative artificial intelligence's representation of social reality. *Poetics*, *108*, 101966..

[18] Klingbeil, A., Grützner, C., & Schreck, P. (2024). Trust and reliance on AI—An experimental study on the extent and costs of overreliance on AI. *Computers in Human Behavior*, *160*, 108352.

[19] Zhang, L., & Yu, L. (2023). From traditional governance to agile governance: A paradigm shift in generative AI governance. *E-Government*, (9), 2-13. (in Chinese)

[20] Grossmann, I., Feinberg, M., Parker, D. C., Christakis, N. A., Tetlock, P. E., & Cunningham, W. A. (2023). AI and the transformation of social science research. *Science*, *380*(6650), 1108-1109.

[21] Pang, X. (2024). AI empowering social science research: Generative agents, complex causal analysis, and human-AI research collaboration. *World Economics and Politics*, (7), 3-28. (in Chinese)

[22] Cai, Z., & Zhang, J. (2024). Value reflection and path exploration of generative AI embedded in digital government. *Journal of Harbin Institute of Technology (Social Sciences Edition),* 26(6), 151-160. (in Chinese)

[23] Li, C., & Yang, J. (2023). Generative AI empowering digital rural governance: Practices, risks, and preventive measures. *Journal of Yunnan Minzu University (Philosophy and Social Sciences Edition)*, 40(6), 107-115. (in Chinese)

[24] Yu, G., & Su, J. (2023). The communication revolution and media ecology in the wave of generative AI: From ChatGPT to the future of comprehensive intelligent era. *Journal of Xinjiang Normal University (Philosophy and Social Sciences Edition),* 44(5), 81-90. (in Chinese)

[25] Lu, Y., Aleta, A., Du, C., Shi, L., & Moreno, Y. (2024). LLMs and generative agent-based models for complex systems research. *Physics of Life Reviews*.(51), 283-293.

[26] Deroy, A., Ghosh, K., & Ghosh, S. (2023). How ready are pre-trained abstractive models and LLMs for legal case judgement summarization?. *arXiv preprint arXiv:2306.01248*.

[27] Nay, J. J., Karamardian, D., Lawsky, S. B., Tao, W., Bhat, M., Jain, R., ... & Kasai, J. (2024). Large language models as tax attorneys: a case study in legal capabilities emergence. *Philosophical Transactions of the Royal Society A*, *382*(2270), 20230159.

[28] Ziems, C., Held, W., Shaikh, O., Chen, J., Zhang, Z., & Yang, D. (2024). Can large language models transform computational social science?. *Computational Linguistics*, *50*(1), 237-291.

[29] Yu, X., & Wang, J. (2024). The "truth" gap: A comparative study of human cognition and large language models in news veracity judgment. *Contemporary Communication*, 2024(5), 17-23. (in Chinese)

[30] Liu, X., Liu, H., Yang, G., Jiang, Z., Cui, S., Zhang, Z., ... & Wang, G. (2025). A generalist medical language model for disease diagnosis assistance. *Nature Medicine*, 1-11.

[31] Cabral, S., Restrepo, D., Kanjee, Z., Wilson, P., Crowe, B., Abdulnour, R. E., & Rodman, A. (2024). Clinical reasoning of a generative artificial intelligence model compared with physicians. *JAMA Internal Medicine*, *184*(5), 581-583.

[32] Yu, M. (2024). Construction logic and implementation path of fertility support policies under "dual objectives". *Social Security Review*, *8*(2), 148-159. (in Chinese)

[33] Holzmann, R., & Wang, X. (2024). Rethinking "multi-pillar pensions". *Social Security Review*, *8*(3), 38-56. (in Chinese)

[34] Zhang, Y. (2024). Efficiency measurement, spatial network characteristics and influencing factors of elderly care institutions under high-quality development. *Social Security Review*, *8*(1), 107-125. (in Chinese)

[35] Zheng, S. (2024). Family care scenarios and policy choices for disabled elderly. *Social Security Review*, *8*(3), 108-126. (in Chinese)

[36] Zhang, S., & Zhang, H. (2024). Utilization analysis and optimization measures of home-based community elderly care facilities in China. *Social Security Review*, *8*(1), 88-106. (in Chinese)

[37] Wang, J. (2024). "Market contracting system" in rural elderly care service provision: Operational model and practical logic. *Social Security Review*, *8*(4), 121-137. (in Chinese)

[38] Zhao, Q. (2024). Social insurance participation choices and influencing factors among gig workers: A survey of delivery riders and couriers. *Social Security Review*, *8*(3), 57-74. (in Chinese)

[38] Sun, S. (2024). Legal dilemmas and reform reconstruction of individual medical accounts in employee basic medical insurance. *Social Security Review*, *8*(4), 86-101. (in Chinese)

[40] Zheng, B., & Wei, W. (2024). 25 years of China's medical payment system reform: Achievements, problems and prospects. *Social Security Review*, *8*(3), 75-89. (in Chinese)

[41] Cai, J. (2024). Path selection for resolving social security dilemmas in flexible employment. *Social Security Review*, *8*(1), 33-45. (in Chinese)